\documentclass[10pt]{amsart} 
\usepackage{amssymb} 
\usepackage{latexsym} 
\usepackage{amsfonts,euscript} 

\usepackage{graphicx} 
\usepackage{verbatim}

\newtheorem{theorem}{Theorem}[section]

\newtheorem{remark}[theorem]{\sc Remark}

\newcommand{\lam}{\lambda} 
\newcommand{\no}{\noindent} 
\newcommand{\e}{\varepsilon}

\def\R{{ \mathbb{R}}}

\def\cal {\EuScript} 
 \def\bC{{\bf C}}

\def\di{\diamond}

\numberwithin{equation}{section} 

\begin{document} 
\bibliographystyle{abbrv}

\title{Immune response to a malaria infection: properties of a mathematical model}

\date{June 29, 2007} 
\begin{abstract} 
We establish some properties of a within host mathematical model of malaria proposed by Recker 
et al \cite{Nature,BMB} which includes the role of the immune system during the infection. The model
accounts for the antigenic variation exhibited by the malaria parasite (P. falciparum).
We show that the model can exhibit a wide variety 
of dynamical behaviors. We provide criteria for global stability, competitive exclusion,
and persistence. We also demonstrate that the disease equilibrium can be destabilized
by non-symmetric cross-reactive responses.
\end{abstract} 
\maketitle 
\author{\begin{center}
Patrick De Leenheer\footnote{email: {\tt deleenhe@math.ufl.edu}. Supported in part by NSF grant DMS-0614651.}
and 
Sergei S. Pilyugin\footnote{email: {\tt pilyugin@math.ufl.edu}. Supported in part by NSF grant DMS-0517954.}, \\
Department of Mathematics, University of Florida, Gainesville, FL 32611-8105, USA.\end{center}
\begin{center}{\it To our mentor and good friend Hal Smith, on the occasion of his $60$th birthday.}\end{center}}

\section{Introduction}

This paper addresses the within host dynamics of the malaria parasite {\it Plasmodium falciparum} whose 
infection mechanism we briefly review here. Infection starts when a human 
is bitten by an infected mosquito that releases sporozoites in the bloodstream. The sporozoites quickly enter the liver where they mature, replicate, and differentiate into merozoites. The  merozoites are then released into the bloodstream, where they go on to infect   
erythrocytes (red blood cells). Merozoites reproduce within infected erythrocytes for a period of about two days. Finally the infected erythrocyte ruptures and releases new merozoites that repeat the infection cycle. A discussion leading to a 
mathematical model that considers a single parasite strain can be found in \cite{saul,gravenor} and references therein. 
In practice however, there is a considerable diversity among the infected erythrocytes, which is reflected by a wide variety of the surface proteins (antigens) that are presented by the infected cells. A mathematical model that includes an arbitrary number of parasite strains was studied in a very elegant paper by Iggidr et al \cite{iggidr} where a competitive exclusion principle was established. Generically, only one strain survives while the others are driven to extinction. 

The mathematical models mentioned above do not include any immune response mounted by the human host. Although many details of immune responses to {\it P. falciparum} are presently not well understood, there is evidence that the antigenic variation between different strains of the
parasite prompts the immune system to mount both strain specific as well as cross reactive responses \cite{antia,Nature,BMB}. The primary distinction between specific and cross-reactive responses is that they target major (unique to each strain) or minor (shared
among strains) epitopes, respectively, on the infected cell's surface.

The goal of this paper is to extend the analysis of the model proposed in \cite{Nature,BMB} which include 
the different immune responses described above. We provide some results concerning the global 
behavior of this model by 
\begin{enumerate} 
\item Showing global asymptotic stability of the system in two extreme cases 
(no cross immunity and perfect cross immunity). 
\item Showing the possibility of oscillatory destabilization in the case of partial cross immunity. 
\item Establishing conditions for both competitive exclusion as well as for persistence. 
\end{enumerate} 
Our results indicate that depending on parameter values, 
this model can exhibit a wide variety of dynamical behaviors. The full range of possible behaviors and 
biological implications is currently not fully understood and remains the objective of future research. 

The rest of this paper is organized as follows. In Section 2 we recall and slightly generalize 
the model from \cite{Nature,BMB} 
and in Section 3 we comment on the existence or non-existence of positive equilibria. In 
Section 4 we treat the case of a single parasitic strain and establish global asymptotic stability,   
even when the growth rate of infected cells is assumed to be logistic, as opposed to linear. Similar 
results are obtained in Section 5 in two special cases: the case of no cross immunity, and the case 
of perfect cross immunity. In case of partial cross immunity, the dynamic picture is not as simple and 
this is illustrated in Section 6 by analyzing a particular example. In Section 7 we return to the general 
model and establish sufficient conditions for competitive exclusion as well as for persistence. These 
conditions are compared to similar ones for certain associated Lotka-Volterra systems of lower dimension. 

\section{General modeling assumptions} 
The model that we study here was originally proposed by 
Recker et al \cite{Nature} and later analyzed by Recker and Gupta \cite{BMB}. 
The model has the following form 
\begin{eqnarray} 
\dot y_i & = & y_i(\phi - \alpha z_i - \alpha' w_i), \label{eqY'}\\ 
\dot z_i & = & \beta y_i - \mu z_i,\label{eqZ'}\\ 
\dot w_i & = & \beta' \sum_{j=1}^n c_{ij} y_j - \mu' w_i, \label{eqW'} 
\end{eqnarray} 
where $i=1,...,n.$ The variables $y_i$, $z_i$, and $w_i$ represent the 
abundance of the erythrocytes which are infected by the $i$-th parasite, 
and the magnitudes of the specific and cross-reactive immune response respectively. 
We assume that 
the immune responses are induced proportionally to the parasitic load at the 
rates $\beta$ and $\beta'$. The coefficients $\mu$ and $\mu'$ model the 
life-span of the corresponding immune responses. The efficiency of both 
responses are given by $\alpha$ and $\alpha'$. The coefficient $\phi$ 
represents the maximal growth rate of the parasite. We assume that all 
kinetic parameters are equal for all strains. Finally, we 
assume that each strain has a distinct major epitope, but two different 
strains may share common minor epitopes. In the model, we incorporate 
this assumption by introducing the non-negative cross-reactivity matrix $C$ such that 
$c_{ij}>0$ if the strains $i$ and $j$ share the same epitope and $c_{ij}=0$ 
otherwise. In the sequel we will refer to some special cases for which we introduce the 
following terminology: 
\begin{enumerate} 
\item We say that there is no cross immunity when $C=I$.   
\item We say that there is perfect cross immunity when $C={\bf 1}'{\bf 1}$, where ${\bf 1}=(1\dots 1)\in \R$. 
\item Otherwise we say that there is partial cross immunity. 
\end{enumerate} 
For mathematical convenience, we perform a simple rescaling of the original variables
and rewrite system $(\ref{eqY'})-(\ref{eqW'})$ as
\begin{eqnarray} 
\dot y_i & = & y_i(1 - z_i - w_i), \label{eqY}\\ 
\dot z_i & = & y_i - \mu_1 z_i,\label{eqZ}\\ 
\dot w_i & = & b \sum_{j=1}^n c_{ij} y_j - \mu_2 w_i, \label{eqW} 
\end{eqnarray}
and define $\gamma_1:=\frac{1}{\mu_1}$
and $\gamma_2:=\frac{b}{\mu_2}.$

In case of P. falciparum, there is a natural carrying capacity given by the 
number of available erythrocytes which can be infected by the parasite. 
Setting aside the possible effects of erythropoesis, we can assume that 
such carrying capacity is constant and modify the model accordingly, 
\begin{eqnarray} 
\dot y_i & = & y_i(1-\frac{y}{K} - z_i -  w_i),\\ 
\dot z_i & = &  y_i - \mu_1 z_i,\\ 
\dot w_i & = & b \sum_{j=1}^n c_{ij}  y_j - \mu_2 w_i.   
\end{eqnarray} 

\section{The positive equilibrium} 
Using the vector notation, we can express the equilibrium conditions of $(\ref{eqY})-(\ref{eqW})$ as follows: 
$ {\bf z}^\ast =\gamma_1 {\bf y}^\ast,$ and
$ {\bf w}^\ast=\gamma_1 C {\bf y}^\ast.$ 
The positive equilibrium must then satisfy the condition 
\begin{equation} \gamma_1  {\bf y}^\ast+\gamma_2 C {\bf y}^\ast={\bf 1}. 
\label{pos}\end{equation} 
In case of perfect cross-reactivity, where $c_{ij}=1$ for all $i,j$, 
there exists a positive solution of the form 
$$ y_i^*=\bar{y}=\frac{1}{\gamma_1 + n \gamma_2}, \quad i=1,..., n,$$ 
which corresponds to a positive equilibrium. 

The positive equilibrium does not always exist. For instance, letting $n=3$, 
$\gamma_1=\gamma_2=1$ and 
\begin{equation} 
C=\left( \begin{array}{ccc}1 & 1+\epsilon & 0\\ 1+\epsilon & 1 & 1+\epsilon \\ 0 & 1+\epsilon & 1  \end{array}\right)
\label{C}\end{equation} 
the solution of (\ref{pos}) is given by 
$ (y_1^*,y_2^*,y_3^*)=1/2(2-(1+\epsilon)^2)\left(1-\epsilon,-2\epsilon,1-\epsilon\right)$, which is non-negative for 
$\epsilon=0$, positive for small negative $\epsilon$, and neither for small positive $\epsilon$. 

\section{Global stability in case $n=1$} 
In the simplest case $n=1$, the model 
\begin{eqnarray} 
\dot y & = & y(1 -  z -  w), \label{eqY1}\\ 
\dot z & = &  y - \mu_1 z,\label{eqZ1}\\ 
\dot w & = & b y - \mu_2 w, \label{eqW1} 
\end{eqnarray} 
admits a unique positive equilibrium 
$$ (y^*,z^*,w^*)=\biggl(\frac{1}{\gamma_1+\gamma_2}, \frac{\gamma_1} 
{\gamma_1+\gamma_2}, 
\frac{\gamma_2}{\gamma_1+\gamma_2} \biggr)$$ 
which is globally stable. To see this, we rewrite (\ref{eqY1}--\ref{eqW1}) 
as 
\begin{eqnarray*} 
\dot y & = & y((z^*-z) + (w^*- w)), \\ 
\dot z & = & (y-y^*) - \mu_1 (z-z^*),\\ 
\dot w & = & b (y-y^*) - \mu_2 (w-w^*), 
\end{eqnarray*} 
and define
$$ V= \int_{y^*}^y \frac{s-y^*}{s}\, ds + \int_{z^*}^z (s-z^*)\, ds 
+ \frac{1}{b}\int_{w^*}^w (s-w^*)\, ds.$$ 
The function $V$ clearly has a unique global minimum at $(y^*,z^*,w^*)$. In addition, 
$$ \dot V = (y-y^*)\bigl((z^*-z) + (w^*- w)\bigr) 
+(z-z^*)\bigl( (y-y^*) - \mu_1 (z-z^*) \bigr)+ 
\frac{1}{b}(w-w^*)\bigl( b (y-y^*) - \mu_2 (w-w^*) \bigr)$$ 
which simplifies to 
$$ \dot V = -\mu_1 (z-z^*)^2 -\frac{\mu_2}{b}(w-w^*)^2.$$ 
Clearly, the equilibrium $(y^*,z^*,w^*)$ is the only invariant set in $\{ \dot V=0\}$. 
LaSalle's invariance principle then implies global stability of $(y^*,z^*,w^*)$. 

Assuming a carrying capacity for the infected cells we have a different model 
\begin{eqnarray} 
\dot y & = & y(1-\frac{y}{K} -  z -  w), \label{eqY2}\\ 
\dot z & = &  y - \mu_1 z,\label{eqZ2}\\ 
\dot w & = & b y - \mu_2 w. \label{eqW2} 
\end{eqnarray} 
It is easy to see that the modified model (\ref{eqY2}--\ref{eqW2}) also 
admits a unique positive equilibrium $(y^*,z^*,w^*)$. Using the same 
function $V$ as before, we observe that 
$$ \dot V = -\frac{1}{K}(y-y^*)^2-\mu_1(z-z^*)^2 -\frac{\mu_2}{b}(w-w^*)^2.$$ 
We conclude again that the positive equilibrium is globally asymptotically stable. 

\section{Global stability in case $n>1$} 
When there are two or more strains present, they can be antigenically 
distinct (no cross-reactivity), or antigenically 
similar (perfect cross-reactivity, see above), or there may be 
partial cross-reactivity. In this section, we prove global convergence 
for the first two cases. We also show that adding a carrying capacity 
does not alter the conclusions. 
\subsection{Perfect cross-reactivity without carrying capacity} 
The equations are 
\begin{eqnarray} 
\dot y_i & = & y_i(1 - z_i - w_i), \label{eqY6}\\ 
\dot z_i & = &  y_i - \mu_1 z_i,\label{eqZ6}\\ 
\dot w_i & = & b \sum_{j=1}^n y_j - \mu_2 w_i, \label{eqW6} 
\end{eqnarray} 
for $i=1,...,n$ and they admit a unique positive equilibrium.   
We observe that for all $i,j$ 
$$\dot w_i -\dot w_j=-\mu_2 (w_i -w_j),$$ 
hence all pairwise differences $w_i-w_j$ decay exponentially to zero. 
To make this argument formal, using $w=w_1$ and $w_j=w+u_j$ for $j\neq 1$ 
we rewrite 
equations (\ref{eqY6}--\ref{eqW6}) as 
\begin{eqnarray} 
\dot y_1&=&  y_1(1 - z_1 -  w),\;\; 
\dot y_j  =  y_j(1 -  z_j - (w+u_j)),\;\; j\neq 1, \label{eqY7}\\ 
\dot z_i & = &  y_i - \mu_1 z_i,\label{eqZ7}\\ 
\dot w & = & b \sum_{j=1}^n y_j - \mu_2 w, \label{eqW7}\\ 
\dot u_j & = & -\mu_2 u_j,\;\; j\neq 1. \label{eqU7} 
\end{eqnarray} 
Clearly, the system (\ref{eqY7}--\ref{eqU7}) is asymptotic to 
the limiting system 
\begin{eqnarray} 
\dot y_i & = & y_i(1 -  z_i -  w), \label{eqY8}\\ 
\dot z_i & = &   y_i - \mu_1 z_i,\label{eqZ8}\\ 
\dot w & = & b \sum_{j=1}^n y_j - \mu_2 w. \label{eqW8} 
\end{eqnarray} 
The Lyapunov function for (\ref{eqY8}--\ref{eqW8}) has the form 
$$ V= \sum_{i=1}^n \biggl(\int_{y_i^*}^{y_i} \frac{s-y_i^*}{s}\, ds + 
\int_{z_i^*}^{z_i} (s-z_i^*)\, ds \biggr) 
+ \frac{1}{b}\int_{w^*}^{w} (s-w^*)\, ds .$$ 
Indeed, after simplifications, we find that 
$$ \dot V= -\mu_1 \sum_{i=1}^n (z_i-z_i^*)^2 -\frac{\mu_2}{b} (w-w^*)^2,$$ 
and then global asymptotic stability follows from Lasalle's invariance principle.
\subsection{Perfect cross-reactivity with carrying capacity} 
The equations are 
\begin{eqnarray} 
\dot y_i & = & y_i\bigl(1-\frac{1}{K}\sum_{j=1}^n y_j - z_i -  w_i \bigr), \label{eqY9}\\ 
\dot z_i & = &  y_i - \mu_1 z_i,\label{eqZ9}\\ 
\dot w_i & = & b \sum_{j=1}^n y_j - \mu_2 w_i, \label{eqW9} 
\end{eqnarray} 
for $i=1,...,n$ and they admit a unique positive equilibrium.   
Arguing as before, we consider the limiting system 
\begin{eqnarray} 
\dot y_i & = & y_i\bigl(1-\frac{1}{K}\sum_{j=1}^n y_j- z_i -  w \bigr), \label{eqY10}\\ 
\dot z_i & = &  y_i - \mu_1 z_i,\label{eqZ10}\\ 
\dot w & = & b \sum_{j=1}^n y_j - \mu_2 w, \label{eqW10} 
\end{eqnarray} 
for which the Lyapunov function is 
$$ V= \sum_{i=1}^n \biggl(\int_{y_i^*}^{y_i} \frac{s-y_i^*}{s}\, ds + 
\int_{z_i^*}^{z_i} (s-z_i^*)\, ds \biggr) 
+ \frac{1}{b}\int_{w^*}^{w} (s-w^*)\, ds .$$ 
Indeed, after simplifications, 
$$ \dot V= -\frac{1}{K} \biggl( \sum_{j=1}^n (y^*_j-y_j) \biggr)^2 
-\mu_1 \sum_{i=1}^n (z_i-z_i^*)^2 -\frac{\mu_2}{b} (w-w^*)^2,$$ 
implying global asymptotic stability of the positive equilibrium.

\section{Analysis of a specific case with $n=2$ and partial cross immunity} 
In this section, we consider the dynamics of the system with $n=2$ 
and $C$ given by 
$$ 
C=\left( \begin{array}{cc}1&1\\2&1\end{array} \right). 
$$ 
Notice that the dynamics of this system also arises when restricting the system with 
$C$ given by (\ref{C}) and $\epsilon=0$ to the invariant set $\{y_1=y_3,z_1=z_3,w_1=w_3\}$. 
The resulting equations have the following form 
\begin{eqnarray*} 
\dot y_i & = & y_i(1-z_i-w_i),\quad i=1,2,\\ 
\dot z_i & = & y_i-\mu_1 z_i, \quad i=1,2,\\ 
\dot w_1 & = & b (y_1+y_2) - \mu_2 w_1,\\ 
\dot w_2 & = & b (2y_1+y_2) - \mu_2 w_2. 
\end{eqnarray*} 
We re-introduce the coefficients $\gamma_1:=1/\mu_1$ and $\gamma_2:=b/\mu_2$. 
The Jacobian of the system is given by 
$$J =\left(\begin{array}{cccccc} 
1-z_1-w_1 & -y_1 & -y_1 & 0 & 0 & 0\cr 
1 & -\mu_1 & 0 & 0 & 0 & 0 \cr 
b & 0 & -\mu_2 & b & 0 & 0\cr 
0 & 0 & 0 & 1-z_2-w_2 & -y_2 & -y_2\cr 
0 & 0 & 0 & 1 & -\mu_1 & 0 \cr 
2b & 0 & 0 & b & 0 & -\mu_2\cr 
\end{array}\right).$$ 

This model admits at most four equilibria: 
\begin{enumerate} 
\item 
The zero equilibrium $E_{00}$ always exists and is always unstable 
since the Jacobian $J(E_{00})$ (not shown)
has eigenvalues $\lam_{1,2}=1,\ \lam_{3,4}=-\mu_1,\ \lam_{5,6}=-\mu_2$. 

\item 
The semitrivial equilibrium 
$$E_{10} =\biggl(\frac{1}{\gamma_1+\gamma_2}, \frac{\gamma_1}{\gamma_1+\gamma_2}, 
\frac{\gamma_2}{\gamma_1+\gamma_2},0,0,\frac{2 \gamma_2}{\gamma_1+\gamma_2}\biggr)$$ 
always exists. The Jacobian $J(E_{10})$ (not shown)
has eigenvalues $\lam_4=\frac{\gamma_1-\gamma_2}{\gamma_1+\gamma_2},\ 
\lam_5=-\mu_1,\ \lam_6=-\mu_2,$ and $\lam_{1,2,3}$ are eigenvalues 
of the matrix 
$$\left(\begin{array}{ccc} 
0 & -y_1 & -y_1\cr 
1 & -\mu_1 & 0 \cr 
b & 0 & -\mu_2 \cr 
\end{array}\right). $$ 
From the preceding stability analysis in Section 3, we already know that $\Re(\lam_{1,2,3}) 
\leq 0$. Using the Routh-Hurwitz criterion, it is not difficult to show that in 
fact $\Re(\lam_{1,2,3})< 0$. Hence, the stability of $E_{10}$ is determined by 
the sign of $\lam_4$. Specifically, $E_{10}$ is (locally) stable if $\gamma_1< 
\gamma_2$, and unstable if $\gamma_1>\gamma_2$. 
\item The semitrivial equilibrium 
$$E_{01} =\biggl(0,0,\frac{\gamma_2}{\gamma_1+\gamma_2},\frac{1}{\gamma_1+\gamma_2}, \frac{\gamma_1}{\gamma_1+\gamma_2}, 
\frac{\gamma_2}{\gamma_1+\gamma_2}\biggr)$$ 
always exists. The Jacobian $J(E_{01})$ (not shown)
has eigenvalues $\lam_1=\frac{\gamma_1}{\gamma_1+\gamma_2},\ 
\lam_2=-\mu_1,\ \lam_3=-\mu_2,$ and $\lam_{4,5,6}$ are eigenvalues 
of the submatrix 
$$\left(\begin{array}{ccc} 
0 & -y_2 & -y_2\cr 
1 & -\mu_1 & 0 \cr 
b & 0 & -\mu_2 \cr 
\end{array}\right). $$ 
As we argued previously, $\Re(\lam_{4,5,6})< 0$. Since $\lam_1>0$, 
$E_{01}$ is always unstable. 
\item The nontrivial equilibrium $E_{11}$ exists if and only if $\gamma_1>\gamma_2$, i.e. precisely when $E_{10}$ 
is unstable. 
The $(y_1,y_2)$ coordinates of $E_{11}$ are given by 
$$ y_1=\frac{\gamma_1}{(\gamma_1+\gamma_2)^2-2\gamma_2^2},\quad 
y_2=\frac{\gamma_1-\gamma_2}{(\gamma_1+\gamma_2)^2-2\gamma_2^2}.$$ 
The common denominator is positive iff $\gamma_1>(\sqrt{2}-1)\gamma_2$, 
and the numerator of $y_2$ is positive iff $\gamma_1>\gamma_2$. 
The Jacobian at $E_{11}$ is given by 
\begin{equation} J(E_{11}) =\left(\begin{array}{cccccc} 
0 & -y_1 & -y_1 & 0 & 0 & 0\cr 
1 & -\mu_1 & 0 & 0 & 0 & 0 \cr 
b & 0 & -\mu_2 & b & 0 & 0\cr 
0 & 0 & 0 & 0 & -y_2 & -y_2\cr 
0 & 0 & 0 & 1 & -\mu_1 & 0 \cr 
2b & 0 & 0 & b & 0 & -\mu_2\cr 
\end{array}\right).\label{JacE11}\end{equation} 
As we showed previously, 
$$\det J(E_{11}) =y_1 y_2 \mu_1^2 \mu_2^2((\gamma_1+\gamma_2)^2-2\gamma_2^2)>0,$$ 
thus $J(E_{11})$ cannot have zero eigenvalues. It  turns out, that in 
the special case $\mu_1=\mu_2=\mu$, all six eigenvalues of $J(E_{11})$ have 
strictly negative real parts: 
If $\mu_1=\mu_2=\mu$, the characteristic polynomial of $J(E_{11})$ 
has the following form: 
$$ p(\lam)=(\mu+\lam)^2 \biggl(\xi^2 + \xi(1+b)(y_1+y_2)+ 
y_1 y_2 (1+2b-b^2) \biggr),$$ 
where $\xi=\lam(\mu+\lam)$. Clearly, two roots are given by $\lam_{1,2}=-\mu$. 
The remaining four roots can be obtained by solving the quadratic equation in $\xi$. 
We have 
$$ y_1=\frac{\mu}{1+2b-b^2}, \quad y_1=\frac{\mu(1-b)}{1+2b-b^2},$$ 
hence $b\in [0,1)$. Substituting the values of $y_1$ and $y_2$, we have 
$$\xi^2 + \xi\frac{\mu(1+b)(2-b)}{1+2b-b^2}+\frac{\mu^2(1-b)}{1+2b-b^2}=0.$$ 
The discriminant of this equation is 
$${\cal D}=\mu^2\frac{(1+b)^2(2-b)^2-4(1-b)(1+2b-b^2)}{(1+2b-b^2)^2}.$$ 
Simplifying the numerator, we find that 
$${\cal D}=\mu^2\frac{b^2(3-b)^2}{(1+2b-b^2)^2} \geq 0.$$ 
Hence the roots are 
$$\xi_1=-\mu, \quad \xi_2=-\mu\frac{1-b}{1+2b-b^2}.$$ 
The corresponding lambdas are solutions of 
$$ \lam_{3,4}^2+\mu \lam_{3,4}+\mu=0, \quad 
\lam_{5,6}^2+\mu \lam_{5,6}+\mu\frac{1-b}{1+2b-b^2}=0.$$ 
The positivity of coefficients in the above quadratics implies that 
$\Re(\lam_{3,4,5,6})<0$. 
\end{enumerate} 

\subsection{Destabilizing the nontrivial equilibrium} 
In this section, we show that there exist a nonempty set of parameter 
combinations such that $E_{11}$ is unstable. To do so, we fix the 
value $b=1$ and let $\mu_1=\e$, $\mu_2=c \e$ where $c>0$ and $\e$ is 
small. Recalculating the equilibrium values, we find 
$$ y_1=\frac{c^2 \e}{c^2+2 c-1}, \quad y_2=\frac{c (c-1) \e}{c^2+2 c-1}.$$ 
The positive equilibrium exists for all $\e>0$ if and only if $c>1$. 
The Jacobian of interest has the form (\ref{JacE11}) with $y_1,y_2,\mu_1,\mu_2$ given above.
\begin{comment}
$$J(\e)=\begin{pmatrix} 
0 & -\frac{c^2 \e}{c^2+2 c-1} & -\frac{c^2 \e}{c^2+2 c-1} & 0 & 0 & 0\cr 
1 & -\e & 0 & 0 & 0 & 0 \cr 
1 & 0 & -c \e & 1 & 0 & 0\cr 
0 & 0 & 0 & 0 & -\frac{c (c-1) \e}{c^2+2 c-1} & -\frac{c (c-1) \e}{c^2+2 c-1}\cr 
0 & 0 & 0 & 1 & -\e & 0 \cr 
2 & 0 & 0 & 1 & 0 & -c \e\cr 
\end{pmatrix}.$$ 
\end{comment}
The characteristic polynomial of $J(\e)$ has the form 
\begin{eqnarray*} p(z,\e) & = & \e^4 a_0(c)+\e^3a_1(c)(1+O(\e))z+\e^2 a_2(c)(1+O(\e))z^2\\ 
& & + 
\e^2 a_3(c)(1+O(\e))z^3+\e a_4(c)(1+O(\e))z^4+\e a_5(c)z^5+z^6,\end{eqnarray*} 
where 
\begin{eqnarray*} 
a_0(c) & = & \frac{c^3(c-1)}{c^2+2 c-1},\\ 
a_1(c) & = & \frac{4 c^4(c-1)}{(c^2+2 c-1)^2},\\ 
a_2(c) & = & \frac{2 c^3(c-1)}{(c^2+2 c-1)^2},\\ 
a_3(c) & = & \frac{3 c(2c-1)(c^3+3c^2+c-1)}{(c^2+2 c-1)^2},\\ 
a_4(c) & = & \frac{2 c(2 c-1)}{c^2+2 c-1},\\ 
a_5(c) & = & 2(c+1). 
\end{eqnarray*} 
Since $p(z,0)=z^6$, $J(0)$ has a zero eigenvalue of multiplicity 6. Now we 
expand the roots of $p$ in powers of $\e$.  First, we evaluate $p(k \e^{\alpha},\e)$ 
and find that the leading terms are 
\begin{eqnarray*} p(k \e^{\alpha},\e) & = & \e^4 a_0(c)+k \e^{3+\alpha} a_1(c)(1+O(\e))+k^2 \e^{2+ 2 \alpha} a_2(c)(1+O(\e))\\ 
& & + 
k^3 \e^{2+3\alpha} a_3(c)(1+O(\e))+k^4 \e^{1+4\alpha} a_4(c)(1+O(\e))+k^5 \e^{1+5 \alpha} a_5(c)+k^6 \e^{6 \alpha},\end{eqnarray*} 
Now we construct the Newton diagram, that is, 
$$ n(\alpha)=\min(4,3+\alpha,2+2\alpha,2+3\alpha,1+4\alpha,1+5\alpha,6\alpha),$$ 
which has two positive vertices at $(1/2,3)$ and $(1,4)$. 
Hence, the leading power of $z$ is either $\alpha=1/2$ or $\alpha=1$. 
\begin{itemize} 
\item Case $\alpha=1$ corresponds to $z=k \e + o(\e)$. To determine the 
value of $k$, we set the leading terms of $p(k \e,\e)$ equal to zero 
and obtain the equation $a_0(c)+k a_1(c)+k^2 a_2(c)=0.$ Simplifying this 
equation, we find that it is equivalent to 
$$ \frac{c^3(c-1)}{(c^2+2 c-1)^2}(2 k^2+4 c k +(c^2+2c-1))=0.$$ 
Since $c>1$, the roots are 
$$ k_{1,2}=-c \pm \frac{c-1}{\sqrt{2}}$$ 
which are both strictly negative. 
\item Case $\alpha=1/2$ corresponds to $z=r \e^{1/2} + l \e + o(\e)$. Expanding 
$p(r \e^{1/2} + l \e ,\e),$ 
we find up to the two lowest orders of $\epsilon$ that
\begin{eqnarray*} 
& & p(r \e^{1/2} + l \e ,\e) = \e^3r^2\Bigl(a_2(c)+a_4(c) r^2+r^4\Bigr)\\ 
& & + r \e^{7/2}\Bigl(a_1(c)+ 2 l a_2(c)+ r^2 a_3(c)+4 r^2 l a_4(c)+r^4 a_5(c) 
+6 r^4 l \Bigr). \end{eqnarray*} 
Setting the $\e^3$ term equal to zero, we find that either $r=0$ (in which case 
we are back to the previous step) or that $r$ satisfies the biquadratic equation 
$$ a_2(c)+a_4(c) r^2+r^4=0,$$ 
which is equivalent to 
$$ 2c^3(c-1)+2c(2c-1)(c^2+2c-1)r^2+(c^2+2c-1)^2r^4=0.$$ 
The discriminant of this equation 
$${\cal D}=4 c^2(c^2+2c-1)^2 \left( (2c-1)^2-2c(c-1)\right)=4 c^2(c^2+2c-1)^2 
\left(c^2+(c-1)^2\right)$$ is clearly positive, 
and both roots 
$$ r^2= 
\frac{c}{c^2+2c-1}\bigl( -(2c-1)\pm\sqrt{(2c-1)^2-2c(c-1)} \bigr)$$ 
are strictly negative. Hence, we have two pairs of pure imaginary 
values for $r$: 
\begin{eqnarray*} 
r_{1,2} & = & \pm i \sqrt{\frac{c\bigl((2c-1)+\sqrt{(2c-1)^2-2c(c-1)}\bigr)}{c^2+2c-1}}, \\ 
r_{3,4} & = & \pm i \sqrt{\frac{c\bigl((2c-1)-\sqrt{(2c-1)^2-2c(c-1)}\bigr)}{c^2+2c-1}}. 
\end{eqnarray*} 
Substituting each pair into the $\e^{7/2}$ term and setting it equal to zero, 
we obtain the corresponding values of $l$: 
\begin{eqnarray*} 
l_1 & = & -\frac{a_1(c)+ r_{1,2}^2 a_3(c)+r_{1,2}^4 a_5(c)}{2 a_2(c)+ 
4 r_{1,2}^2 a_4(c)+6 r_{1,2}^4}, \\ 
l_2 & = & -\frac{a_1(c)+ r_{3,4}^2 a_3(c)+r_{3,4}^4 a_5(c)}{2 a_2(c)+ 
4 r_{3,4}^2 a_4(c)+6 r_{3,4}^4}. 
\end{eqnarray*} 
\end{itemize} 
At this point, we have established the existence of six distinct branches of 
eigenvalues for small $\e>0$: 
\begin{eqnarray*} 
z_1 & = & k_1 \e+o(\e),\\ 
z_2 & = & k_2\e +o(\e),\\ 
z_{3,4} & = & l_1 \e + r_{1,2} \e^{1/2}+o(\e),\\ 
z_{5,6} & = & l_2 \e + r_{3,4} \e^{1/2}+o(\e). 
\end{eqnarray*} 
The first two eigenvalues are real and negative for small $\e>0$, so it remains 
to show that either $l_1$ or $l_2$ may be positive for some values of $c$. 

The sign of the  expression $ 2a_2+4 a_4 r^2+6  r^4$ 
can be determined as follows. Consider a cubic polynomial $f(x)=2x(a_2+a_4x+x^2)$ 
which has three simple zeros at 
$r^2_{1,2}<r^2_{3,4}<0$. Since $f(x)>0$ for 
$x>0$, we have that $f'(r^2_{1,2}), f'(0)>0$, and $f'(r^2_{3,4})<0$. 
Thus 
\begin{eqnarray*} 
f'(r^2_{1,2}) & = & 2 a_2(c)+  4 r_{1,2}^2 a_4(c)+6 r_{1,2}^4>0,\\ 
f'(r^2_{3,4}) & = & 2 a_2(c)+ 4 r_{3,4}^2 a_4(c)+6 r_{3,4}^4<0. 
\end{eqnarray*} 

Since the denominators of $l_1$ and $l_2$ have opposite signs, it suffices to 
show that the numerators have the same sign. That would imply that one of $l_i$ 
is positive. We claim that the numerators of $l_1$ and $l_2$ are strictly 
positive for all sufficiently large $c$. Indeed, lets investigate the 
asymptotic behavior of the roots of the quadratics 
$Q_1(x)=a_1(c)+ a_3(c)x+ a_5(c)x^2$ and $Q_2(x)=a_2(c)+ a_4(c)x+ x^2$. 
\begin{itemize} 
\item Equation $Q_1=0$ is equivalent (after dividing through by $2c$) to 
$$ \frac{2 c^3(c-1)}{(c^2+2 c-1)^2}+ \frac{3 (c-1/2)(c^3+3c^2+c-1)}{(c^2+2 c-1)^2} x 
+(1+1/c)x^2=0.$$ 
As $c\to\infty$, the roots of this equation converge to the roots of $2+3x+x^2=0$, 
that is, $x=-2$ or $x=-1$. This follows from the continuity of roots. 
\item Similarly, as $c\to\infty$, the roots of $Q_2=0$ converge to the 
roots of $2+4x+x^2=0$, that is, $x=-2\pm \sqrt{2}$. An equivalent statement 
is that 
$$ \lim_{c\to\infty} r^2_{1,2}=-2-\sqrt{2}, \quad \lim_{c\to\infty} r^2_{3,4}= 
-2+\sqrt{2}.$$ 
\end{itemize} 

Since $-2-\sqrt{2}<-2<-1<-2+\sqrt{2}$ (i.e. the roots of $Q_1$ are 
located between the roots of $Q_2$), we conclude that the numerators of 
$l_1$ and $l_2$ are strictly positive for all sufficiently large values of $c$. 
Since the denominator of $l_1$ (respectively $l_2$) is positive(respectively negative), . 
we conclude that $l_1>0$ and $l_2<0$ for all sufficiently large $c$. 
(Numerically, this happens as long as $c>2.46$.) We summarize the results 
of this section in the following Lemma. 

{\bf Lemma 1.} {\it Let $b=1,\ \mu_1=\e, \ \mu_2=c\e$, and 
$$C=\left(\begin{array}{cc} 1 & 1\cr 2 & 1\cr\end{array}\right),$$ 
then there exist $\e^*>0$ and $c^*>1$ such that for all 
$0<\e<\e^*$ and $c>c^*$, the Jacobian at the positive equilibrium 
$E_{11}$ has two real negative eigenvalues, and two pairs of complex 
eigenvalues with positive and negative real parts respectively. 
In particular, the equilibrium $E_{11}$ is locally unstable with 
two-dimensional unstable manifold. }

\section{Results on boundedness of solutions, competitive exclusion and persistence}

\subsection{Boundedness of solutions} 
Without loss of generality, consider the scaled model 
\begin{eqnarray} 
\label{mal1}\dot y_i & = & y_i(1- z_i - w_i), \\ 
\label{mal2}\dot z_i & = &  y_i - \mu_1 z_i,\\ 
\label{mal3}\dot w_i & = & b \sum_{j=1}^n c_{ij} y_j - \mu_2 w_i, \end{eqnarray} 
and suppose that $b,\mu_1,\mu_2>0$ and $c_{ii}>0$  for all $i$. 

{\bf Theorem 1} {\it All nonnegative solutions of $(\ref{mal1})-(\ref{mal3})$ are ultimately uniformly bounded.} 

{\bf Proof}. 
Without loss of generality, we may consider only positive solutions, that is 
$ y_i(t),z_i(t),w_i(t)>0$. First, it is clear that since $\dot y_i \leq y_i$, we have 
$y_i(t)\leq y_i(0) e^{t}$. Hence, all solutions are defined for $t\geq 0$. Next, 
we introduce the quantities $\alpha_i =y_i/(z_i+w_i)>0$. It follows that 
$$ \dot \alpha_i=\frac{y_i(1- z_i - w_i)(z_i+w_i)-y_i(y_i+b \sum_{j=1}^n c_{ij} y_j 
- \mu_1 z_i- \mu_2 w_i)}{(z_i+w_i)^2}.$$ 
Clearly, this implies that 
$$ \dot  \alpha_i \leq \alpha_i(1-(1+b c_{ii})\alpha_i+ 
\frac{\mu_1 z_i+\mu_2 w_i}{z_i+w_i}).$$ 
Using the fact that 
$$\frac{\mu_1 z_i+\mu_2 w_i}{z_i+w_i} \leq \max(\mu_1,\mu_2), \quad 
z_i,w_i>0, $$ 
we obtain the inequality 
$$ \dot  \alpha_i \leq \alpha_i(1+ \max(\mu_1,\mu_2)-(1+b c_{ii})\alpha_i).$$ 
Hence, $\dot \alpha_i<0$ as long as 
$\alpha_i>\alpha_i^*:=\frac{1+\max(\mu_1,\mu_2)}{1+bc_{ii}}.$ 
Consequently, 
$\alpha_i(t) \leq \hat \alpha_i:=\max(\alpha_i(0),\alpha_i^*)$ for all $t\geq 0$. 
Equivalently, we have that $y_i(t) \leq \hat \alpha_i(z_i(t)+w_i(t))$, which 
implies that 
$$ \dot y_i \leq y_i(1-\frac{y_i}{\hat \alpha_i}), \quad t\geq 0.$$ 
Therefore, $y_i(t)$ is bounded for all $t\geq 0$. Finally, we have that 
$$ \limsup_{t\to\infty} \alpha_i(t) \leq \alpha^*_i, 
\quad  \limsup_{t\to\infty} y_i(t) \leq \alpha^*_i, 
\quad \limsup_{t\to\infty} z_i(t) \leq \frac{\alpha^*_i}{\mu_1}, 
\quad \limsup_{t\to\infty} w_i(t) \leq \frac{b \sum_{j} c_{ij}\alpha^*_j}{\mu_2}. \quad \di $$ 

\subsection{Competitive exclusion} 
Let $\gamma_1=1/\mu_1$ and $\gamma_2=b/\mu_2$, and define 
$A=\gamma_1 I + \gamma_2 C$. 

\medskip{\bf Theorem 2.} {\it Suppose that the following condition 
holds: 
\begin{equation} 
\exists r\in \{1,...,n\}: \forall {\bf x} \geq {\bf 0},\ A{\bf x}\geq {\bf 1} 
\Rightarrow (A{\bf x})_r >1,\label{excl} 
\end{equation} 
then for any positive solution $ y_i(t),z_i(t),w_i(t)>0$ of $(\ref{mal1})-(\ref{mal2})$, we have 
$\lim_{t\to\infty} y_r(t)=0.$} 

In (\ref{excl}), the vector inequalities correspond to the order induced by the 
standard cone $R^n_+$. 
\no{\bf Proof.} Let 
$\langle f(t)\rangle=\frac{1}{t}\int_0^t f(s)\, ds$ 
denote the time-average of the function $f(t)$. Then for any positive 
solution, we have that 
\begin{eqnarray*} 
\langle \dot y_i/y_i \rangle & = & 1- \langle z_i \rangle - \langle w_i \rangle, \\ 
\langle \dot z_i \rangle & = &  \langle y_i \rangle - \mu_1 \langle z_i \rangle,\\ 
\langle \dot w_i \rangle & = & b \sum_{j=1}^n c_{ij} \langle y_j \rangle - \mu_2 
\langle w_i \rangle, \end{eqnarray*} 
Boundedness of solutions implies  that 
\begin{eqnarray*} 
\langle \dot z_i(t) \rangle & =& \frac{z_i(t)-z_i(0)}{t} \to 0, \quad t\to\infty,\\ 
\langle \dot w_i(t) \rangle & =& \frac{w_i(t)-w_i(0)}{t} \to 0, \quad t\to\infty,\\ 
\limsup_{t\to\infty} 
\langle \dot y_i/y_i \rangle & =& 
\limsup_{t\to\infty} 1- \langle z_i(t) \rangle-\langle w_i(t) \rangle\leq 0. 
\end{eqnarray*} 
Without loss of generality, there exists a convex compact set $K \subset R^n_+$ 
such that ${\bf y}(t) \in K$ for all $t\geq 0$. The convexity of $K$ 
implies that $\langle{\bf y}(t)\rangle \in K$ for all $t\geq 0$. 
Let $K'$ be the compact set 
$$ K'=\{{\bf x}\in K:\ A{\bf x} \geq {\bf 1} \}.$$ 
By $(\ref{excl})$, compactness of $K'$ and continuity, 
there exists $\varepsilon>0$ such that $(A{\bf x})_r >1+\varepsilon$ 
for all ${\bf x}\in K'$. Also by continuity, there exists $\delta>0$ such that 
$(A{\bf x})_r >1+\varepsilon/2$ 
for all ${\bf x}$ in the $\delta$-neighborhood of $K'$. 

Now we analyze the averages more carefully. Since 
$$ |\langle z_i \rangle - \gamma_1 \langle y_i \rangle    |\to 0, \quad 
|\langle w_i \rangle - \gamma_2 \sum_{j=1}^n c_{ij} \langle y_j \rangle    | 
\to 0,$$ 
we have that 
$$ \limsup_{t\to\infty} 1- \langle z_i(t) \rangle-\langle w_i(t) \rangle = 
\limsup_{t\to\infty} 1-\gamma_1\langle y_i(t) \rangle 
- \gamma_2 \sum_{j=1}^n c_{ij} \langle y_j(t) \rangle \leq 0, $$ 
that is, 
$$ \liminf_{t\to\infty} (A\langle {\bf y}(t) \rangle)_i \geq 1$$ 
for all $i=1,...,n$. It follows that there exists $T>0$ such that 
${\rm dist}(\langle {\bf y}(t) \rangle,K')<\delta$ for all $t>T$. 
Therefore, 
$(A\langle {\bf y}(t) \rangle)_r >1+\varepsilon/2$ for all $t>T$. 
This in turn implies that there exists $T'>0$ such that 
$$\langle \dot y_r(t)/y_r(t) \rangle=1- \langle z_r(t) \rangle - \langle w_r(t) \rangle < -\varepsilon/4, \quad t >T',$$ 
or equivalently, 
$$ y_r(t) <y_r(0) \exp( -\varepsilon t/4), \quad t >T'.$$ 
This clearly implies that $\lim_{t\to\infty} y_r(t)=0.$ \hfill $\di$ 

\subsection{Partial persistence} 
Let 
\begin{eqnarray} 
\label{sys1} {\dot x}&=&f(x,y)\\ 
\label{sys2} {\dot y}&=&g(x,y) 
\end{eqnarray} 
be a forward complete system on $X\times Y:=\R^n_+\times\R^m_+$. We say that 
$(\ref{sys1})-(\ref{sys2})$ is {\it x-partially (strongly uniformly) persistent} if there is some $\delta>0$ 
so that for all $(x,y)\in \textrm{int}(\R^n_+)\times \textrm{int}(\R^m_+)$ there holds that 
$$ 
\liminf_{t\rightarrow \infty}x_i(t)\geq \delta,\;\; i=1,\dots, n. 
$$ 

Inspired by the persistence result in \cite{hofbauer} we have 

{\bf Theorem 3} {\it Assume that $\partial X \times Y$ is forward invariant for $(\ref{sys1})-(\ref{sys2})$, 
and suppose $K\subset X\times Y$ is a compact 
absorbing set (thus every forward solution of $(\ref{sys1})-(\ref{sys2})$ eventually enters and remains in $K$). 
Let $P:X\times Y\rightarrow \R$ be continuously differentiable and the restriction of $P$ to $\partial X\times Y$ be $0$, and 
positive elsewhere. Assume that there is a continuous function $\psi:X\times Y\rightarrow \R$ so that 
\begin{equation}\label{log-derivative} 
\frac{{\dot P}}{P}=\psi \textrm{ on }X\times Y\setminus (\partial X\times Y) 
\end{equation} 
If for all $(x,y)\in \partial X \times Y$, there is some $T>0$ such that: 
\begin{equation}\label{increase} 
\langle \psi(x(T),y(T))\rangle >0, 
\end{equation} 
then $(\ref{sys1})-(\ref{sys2})$ is $x$-partially persistent.} \\ 
The proof can be found in \cite{hiv-mutations} and is omitted here. 

\begin{remark}\label{een} 
Note that a result similar to Theorem 12.2.2 in \cite{hofbauer}, but now for system $(\ref{sys1})-(\ref{sys2})$, 
remains valid. 
It states that Theorem 3 remains true if 
condition $(\ref{increase})$ holds just for $(x,y)$ which are 
$\omega$ limit points of orbits in $\partial X\times Y$. 
The proof is exactly the same as in \cite{hofbauer}.   
\end{remark} 

We will apply Theorem 3 to prove a persistence result for the malaria model $(\ref{mal1})-(\ref{mal3})$, 
which we re-write in a more compact form first: 
\begin{eqnarray} 
\label{mala1}{\dot X}&=&\textrm{diag}(X)[{\bf 1}-(I_n \;\; I_n)Y],\label{comp1}\\ 
\label{mala2}{\dot Y}&=&-\textrm{diag}({\bf \mu})Y+BX ,\label{comp2} 
\end{eqnarray} 
where $\begin{pmatrix} X\\ Y\end{pmatrix}\in \R^n_+\times \R^{2n}_+$, 
${\bf 1}=(1\dots 1)'\in \R^n$, ${\bf \mu}=(\mu_1 \dots \mu_1 \;\;\mu_2 \dots \mu_2)'\in \R^{2n}$ and 
$$ 
B=\begin{pmatrix}I\\bC \end{pmatrix}. 
$$ 
Note that $\partial \R^n_+\times \R^{2n}_+$ is forward invariant, and that there is a compact absorbing set 
$K$ in $\R^n_+\times \R^{2n}_+$ by Theorem 1. Let 
$$ 
A=(I_n\;\; I_n)\textrm{diag}^{-1}({\bf \mu})B. 
$$ 
We will show the following: 
{\bf Theorem 4} {\it If there is some $p\in\textrm{int}(\R^n_+)$ so that 
\begin{equation}\label{cond} 
p'[{\bf 1} -A{\bar X}]>0, 
\end{equation} 
for all ${\bar X}$ for which $\begin{pmatrix}{\bar X} \\ \textrm{diag}^{-1}({\bf \mu})B{\bar X}\end{pmatrix}$ 
is an equilibrium of $(\ref{mala1})-(\ref{mala2})$ in $\partial \R^n_+ \times \R^{2n}_+$, 
then system $(\ref{mala1})-(\ref{mala2})$ is persistent. }

{\bf Proof}. 
The proof proceeds in two steps. We will first show that system $(\ref{mala1})-(\ref{mala2})$ is 
$X$-partially persistent using Theorem $3$ and Remark \ref{een}. Then we will show that the 
system $(\ref{mala1})-(\ref{mala2})$ is persistent. 

{\it Step 1}. Let us first establish $X$-partial persistence for $(\ref{mala1})-(\ref{mala2})$. 
Define the continuously differentiable (perhaps by multiplying the vector $p$ by a sufficiently 
large positive scalar) function $P:\R^n_+\times \R^{2n}_+ \rightarrow [0,\infty)$: 
$$ 
P(X,Y)=\Pi_{i=1}^nX_i^{p_i}, 
$$ 
which is $0$ on $\partial \R^n_+\times \R^{2n}_+$ and positive elsewhere. Note that 
$(\ref{log-derivative})$ holds on $\R^n_+\times \R^{2n}_+\setminus (\partial \R^n_+\times \R^{2n}_+)$ with 
$$ 
\psi(X,Y)=p'[{\bf 1}-(I_n\;\; I_n)Y] 
$$ 
We claim that for all $Z=(X, Y)\in \partial \R^n_+\times \R^{2n}$, there is some $T>0$ such that: 
$$ 
\langle \psi(Z(T))\rangle >0,
 $$ 
from which $X$-partial persistence will follow using Theorem 3. 
We will do this by induction on $r$, the number of non-zero components of $X$. 
If $r=0$, then $X(t)=0$ for all $t\geq 0$, hence $Y(t)\rightarrow 0$ as $t\rightarrow +\infty$, so that 
$\omega(Z)=\{0\}$. But since $0$ is an equilibrium point of $(\ref{mala1})-(\ref{mala2})$, 
$(\ref{cond})$ holds with ${\bar X}=0$, and therefore our claim follows from Remark \ref{een}. 
Assume that the claim has been established for $r=1,\dots, m-1$ but that $X$ has $m$ non-zero components 
(of course, $m<n$). 
Denote the indices of these components by $I$, a proper subset of $\{0,1,\dots,n\}$. There are two cases to consider: 

{\it Case 1}. The solution $Z(t)$ converges to the boundary of the set 
$D=\{(X\;\; Y)\in \R^n_+\times \R^{2n}_+\; |\;\; X_i\neq 0\textrm{ for all } i\in I\}$. Then $\omega (Z)$ is 
contained in part of the boundary of $\R^n_+\times \R^{2n}_+$ where at most $m-1$ components of $X$ are non-zero. 
The conclusion of our claim then follows from Remark \ref{een} and the induction hypothesis. 

{\it Case 2}. The solution $Z(t)$ does not converge to the boundary of $D$. 
Then there is some $\epsilon>0$ and an increasing  sequence $t_k\rightarrow \infty$ 
so that $X_i(t_k)>\epsilon$ for all $k$ and all $i\in I$. For $i\notin I$ we have that $X_i(t)=0$ for all $t\geq 0$ 
and thus in particular for all $t=t_k$. 
Consider the (bounded) sequences of averages 
$\langle X(t_k)\rangle$ and $\langle Y(t_k) \rangle$, which we may assume -by passing to a subsequence if necessary- 
converge to limits ${\tilde X}$ and ${\tilde Y}$ with the property that ${\tilde X}_i>0$ if $i\in I$ and 
${\tilde X}_i=0$ otherwise. 
Integrating $(\ref{mala2})$ between $0$ and $t_k$, dividing by $t_k$ and letting $t_k\rightarrow \infty$ yields: 
\begin{equation}\label{previous} 
0=-\textrm{diag}{\tilde Y}+B{\tilde X}. 
\end{equation} 
Consider now the dynamics of the components $X_i$ with $i\in I$ as described by $(\ref{mala1})$. 
In particular, dividing by $X_i$, integrating between $0$ and $t_k$, 
dividing by $t_k$ and letting $t_k\rightarrow \infty$, and using $(\ref{previous})$ yields: 
$$ 
0=1-(A{\tilde X})_i,\;\; i\in I. 
$$ 
Since ${\tilde X}_i=0$ for all $i\notin I$ we see that 
$\begin{pmatrix}{\tilde X}\\\textrm{diag}^{-1}(\mu)B{\tilde X}\end{pmatrix}$ is an equilibrium of 
$(\ref{mala1})-(\ref{mala2})$. Finally notice that as $t_k\rightarrow \infty$, we have that: 
$$ 
\langle \psi(Z(t_k))\rangle \rightarrow p'[{\bf 1}-A{\tilde X}], 
$$ 
which is positive by $(\ref{cond})$. This establishes our claim. 

{\it Step 2}. 
In Step 1 we have shown that $(\ref{mala1})-(\ref{mala2})$ is $X$-partially persistent, so that 
for all solutions starting in $\textrm{int}(\R^n_+)\times\textrm{int}(\R^{2n}_+)$ there is some $\delta>0$ such   
that 
$$ 
\liminf_{t\rightarrow \infty}X(t)\geq \delta {\bf 1}, 
$$ 
where the above vector inequality should be interpreted componentwise. 
Then $(\ref{mala2})$ implies that for all large $t$, we have that 
$$ 
{\dot Y}\geq -\textrm{diag}(\mu)Y+\frac{\delta}{2}B{\bf 1}. 
$$ 
This implies that: 
$$ 
\liminf_{t\rightarrow \infty}Y(t)\geq \frac{\delta}{2}\textrm{diag}^{-1}(\mu)B{\bf 1}, 
$$ 
where the vector on the right-hand side has positive components, which establishes persistence of 
$(\ref{mala1})-(\ref{mala2})$. \hfill $\di$

\subsection{Discussion} 
It is interesting to compare our competitive exclusion result (Theorem 2) and our 
persistence result (Theorem 4) obtained in 
the previous subsections to corresponding results for the following lower dimensional Lotka-Volterra system: 
\begin{equation}\label{VL} 
{\dot X}=\textrm{diag}(X)[{\bf 1}-AX] 
\end{equation} 
For this system we can easily prove the following competitive exclusion result, using similar arguments as 
those in the proof of Theorem 2. 

{\bf Lemma 3} {\it Suppose that $(\ref{excl})$ holds for system $(\ref{VL})$. Then for any solution 
$x(t)$ of $(\ref{VL})$ in $\textrm{int}(\R^n_+)$, there holds that $x_r(t)\rightarrow 0$ as $t\rightarrow \infty$.} 

For system $(\ref{VL})$, there is the following persistence result \cite{hofbauer}. 

{\bf Lemma 4} {\it If there is some $p\in \textrm{int}(\R^n_+)$ such that $(\ref{cond})$ holds 
for all ${\bar X}$ which are equilibria of $(\ref{VL})$ in $\partial \R^n_+$, then 
system $(\ref{VL})$ is persistent.} 

In other words, our conditions under which system $(\ref{comp1})-(\ref{comp2})$ exhibits competitive 
exclusion (see Theorem $2$), respectively persistence (see Theorem $4$) holds, 
are the same as for the reduced order system $(\ref{VL})$. 

Finally, we can interpret conditions $(\ref{excl})$ and $(\ref{cond})$ geometrically, and will see that 
they are not mutually exclusive. This implies that there are examples of system $(\ref{comp1})-(\ref{comp2})$ 
which don't fit our conditions for either competitive exclusion or persistence. 

In $\R^n$, define the closed convex set 
$$ 
D=\{x\in \R^n\,|\, {\bf 1}-Ax\leq 0\}. 
$$ 
The boundary of $D$ is given by those points $x$ in $D$ for which $1-(Ax)_i=0$ for some $i$. 
In this case we say that constraint $i$ is active for $x$. 
Condition $(\ref{excl})$ says that there must be a constraint $r$ which is never active in $\R^n_+$. 

Although a geometric interpretation of condition $(\ref{cond})$ is not immediately clear, it has been shown 
in \cite{hofbauer} that $(\ref{cond})$ is equivalent to the following condition 
which does have a clear geometric meaning. 
\begin{equation}\label{equiv} 
C\cap D_+ =\emptyset, 
\end{equation} 
where $C$ is the convex hull of the set of equilibria of $(\ref{VL})$ in $\partial \R^n_+$ and 
$D_+=D\cap \R^n_+$. 

\begin{figure}\label{bis} 
\centering 
\includegraphics[width=6cm]{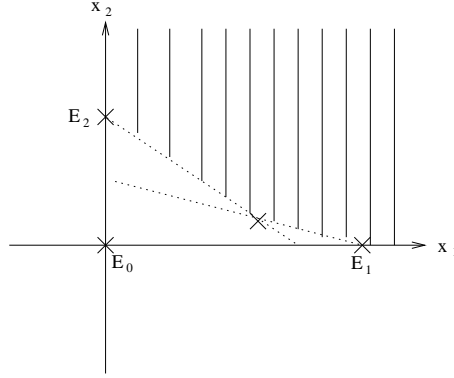} 
\caption{An example of system $(\ref{VL})$ with $n=2$. Nullclines are the dashed lines. 
The hatched region represents $D_+$. The crosses represent the equilibria, 
the triangle $E_0-E_1-E_2$ represents $C$, hence $C \cap D_+\neq\emptyset$.} 
\end{figure}

To see that the exclusion condition $(\ref{excl})$ and $(\ref{cond})$ (or the equivalent $(\ref{equiv})$) 
are not mutually exclusive, consider 
a system $(\ref{VL})$ with $n=2$ with nullclines given in figure $\ref{bis}$   
Clearly neither condition $(\ref{excl})$ nor condition $(\ref{equiv})$ hold. It is well-known that this 
is an example of a bistable Lotka-Volterra system. The equilibrium in $\textrm{int}(\R^2_+)$ is a saddle and every 
solution in $\textrm{int}(\R^2_+)$ not on the stable manifold of the interior equilibrium converges to 
either $E_1$ or $E_2$. 

\end{document}